\documentclass[reprint,showpacs,aps,pre,preprintnumbers]{revtex4-1}
\usepackage{graphicx}
\usepackage{tipa}
\begin{document}

\title{Abstract models for heat engines}
\author{Z. C. Tu}\email[Email: ]{tuzc@bnu.edu.cn}
\affiliation{Department of Physics, Beijing Normal University, Beijing 100875, China}

\date{\today}

\begin{abstract}
We retrospect three abstract models for heat engines which include a classic abstract model in textbook of thermal physics, a primary abstract model for finite-time heat engines, and a refined abstract model for finite-time heat engines. The detailed models of heat engines in literature of finite-time thermodynamics may be mapped into the refined abstract model. The future developments based on the refined abstract model are also surveyed.
\end{abstract}
\preprint{Frontiers of Physics \hspace{3cm}}
\maketitle


\section{Introduction}

Thermodynamics was regarded as a universal theory by Einstein: ``It is the only physical theory of universal content, which I am convinced, that within the framework of applicability of its basic concepts will never be overthrown." The first law (energy conservation) and the second law (entropy increase) are in the theoretical cores of thermodynamics. The whole theoretical system of thermodynamics is developed around these two laws, particularly the second law.

Heat engines, as model systems, demonstrate perfectly the first law and the second law of thermodynamics.
A heat engine undergoing the Carnot cycle which consists of two isothermal processes and two adiabatic processes can achieve the so called Carnot efficiency when these processes are reversible, i.e., quasi-static and frictionless. If one of these processes is irreversible, the efficiency of the engine is smaller than the Carnot efficiency. In other words, the Carnot efficiency is the upper bound of efficiencies for all engines operating between two thermal reservoirs for given temperatures. This is a natural consequence of the first law and the second law when they are applied in heat engines. While the Carnot efficiency is the upper bound, a heat engine undergoing the Carnot cycle results in a vanishing power. This kind of heat engines are useless in practice. The thermodynamic cycle should be speeded up to produce a finite power. The investigations of the efficiency at certain power (in particular maximum power) give birth to finite-time thermodynamics~\cite{Chambadal,Novikov,Curzon1975,Andresen77}. This research field has attracted much attention~\cite{Hoffmann85,devos85,Chen1989,ChenJC94,Bejan96,vdbrk2005,dcisbj2007,Esposito2010,Schmiedl2008,Tu2008,GaveauPRL10,Esposito41106,WangTu2011,ChenLGJNEQ2011,wangtu2012,wangtuctp2013,Izumida2012,wangheinter,Guochen2013,Apertet12a,ApertetPRE13,
AnguloBrown13,QuanHTPRE2014,calvo2015EPJST,IzumidaNJP2015,LongLiuPRE2016,KoningIndekeu2016,YuChenLG2016,ApertetPRE2017,WangPRE96012152,LeeUmPRE052137,MaYHPRE98022133,MedinaPRL2020} for many years.

There are three notable parameter-independent results on the issue of efficiency at maximum power in finite-time thermodynamics. The first one is the Curzon-Ahlborn efficiency,
\begin{equation}\label{eq-CA}
\eta_{1}= 1-\sqrt{1-\eta_0}=\frac{\eta_0}{2} + \frac{\eta_0^2}{8}+ \frac{6\eta_0^3}{96} + O(\eta_0^4),
\end{equation}
the efficiency at maximum power for a macroscopically endoreversible heat engine with Newton-type heat transfer~\cite{Curzon1975}. The notation $\eta_0$ in Eq.~(\ref{eq-CA}) represents the Carnot efficiency. The second one is the Schmiedl-Seifert efficiency,
\begin{equation}\label{eq-SS}
\eta_{2}=\frac{2\eta_0}{4-\eta_0}=\frac{\eta_0}{2} + \frac{\eta_0^2}{8}+
\frac{3\eta_0^3}{96} + O(\eta_0^4),
\end{equation}
the efficiency at maximum power for a Brownian stochastic engine~\cite{Schmiedl2008}. The thermodynamic cycle of this kind of engines is realized by using a time-dependent harmonic potential to control a Brownian particle. We note that the above result happen to be a special case obtained by Chen and Yan for a macroscopically endoreversible heat engine with Fourier-type heat transfer~\cite{Chen1989}. The third one is
\begin{equation}\label{eq-Tuzc}
\eta_3 =\frac{\eta_0^2 }{\eta_0-(1-\eta_0)\ln(1-\eta_0)}=\frac{\eta_0}{2} + \frac{\eta_0^2}{8}+ \frac{7\eta_0^3}{96} + O(\eta_0^4),
\end{equation}
the efficiency at maximum power derived by Tu for the Feynman ratchet as a heat engine~\cite{Tu2008}.

The above three equations coincide up to the quadratic terms of $\eta_0$, which enlightens Tu to conjecture that a universal efficiency at maximum power,
\begin{equation}\label{eq-univ}
\eta_U =\frac{\eta_0}{2} + \frac{\eta_0^2}{8} + O(\eta_0^3),
\end{equation}
should exist for a broad class of heat engines operating between two thermal reservoirs with small temperature differences~\cite{Tu2008}. This universality has been confirmed by a quantum-dot engine~\cite{Esposito2009a}. The universality up to the linear term of Eq.~(\ref{eq-univ}) was found by Van den Broeck~\cite{vdbrk2005} for tight-coupling heat engines working at maximum power. Then, considering a process of particle transports, Esposito \emph{et al.} demonstrated that the universality of the quadratic term in Eq.~(\ref{eq-univ}) comes from left-right symmetry for a tight-coupling engine~\cite{Esposito2009PRL}. This finding was confirmed by other nonlinear models of heat engines~\cite{Izumida2012,ApertetPRE13,ShengTuJPA13}.
Unfortunately, an apparent paradox is that the Curzon-Ahlborn endoreversible engine and the Feynman ratchet still recover universal efficiency (\ref{eq-univ}) in the absence of symmetry. The efficiency at maximum power for the Curzon-Ahlborn heat engine is irrelevant to specific model-dependent parameters, and so regardless of any symmetry. In the extremely asymmetric case, the efficiency at maximum power for the Feynman ratchet still equates universal form (\ref{eq-univ}). Seifert also pointed out that the Feynman ratchet still holds the universality in other asymmetric cases~\cite{Seifert12rev}. This paradox implies that the demonstration by Esposito \emph{et al.} is not a full proof to the conjecture of universal efficiency at maximum power.

There are two gaps in the proof to the conjecture of universal efficiency at maximum power. First, we need a generic abstract model rather than specific models for finite-time (or finite-rate) heat engines when we discuss the universality. Second, we need a nonlinear relation for response because the universal form (\ref{eq-univ}) contains the second order term of $\eta_0$. Sheng and Tu overcame the above two obstacles and then resolved the paradox in a series of work~\cite{ShengTuPRE14,ShengTuPRE15}. By introducing new concepts of weighted reciprocal of temperature and weighted thermal flux, they proposed a generic refined abstract model for finite-time heat engines~\cite{ShengTuPRE14}. In this review, we will revisit the abstract model and its related topics. The rest of this review is organized as follows. In Sec.~\ref{sec-thermodyn}, we present a bird's-eye view of classic thermodynamics. The classic abstract model for heat engines and the primary abstract model for finite-time heat engines
are illustrated. In Se.~\ref{sec-refinedmodel}, we present a refined abstract model for finite-time heat engines and discuss the issue of universal efficiency at maximum power. In Sec.~\ref{sec-appl}, we map the detailed models of heat engines in literature into the refined abstract model. In Sec.~\ref{sec-summary}, we discuss the possible developments in the future on the basis of the refined abstract model.

\section{Bird's-eye view of Thermodynamics\label{sec-thermodyn}}
In this section, we will sketch the outline of thermodynamics. We particularly concern entropy production and the second law of thermodynamics. The rate of entropy production may expressed as a canonical form, the sum of products of thermodynamic fluxes and forces. The key ingredient of linear irreversible thermodynamics is the Onsager reciprocal relation~\cite{Onsager1931a,Casimir1945}. Within the framework of linear irreversible thermodynamics and with the consideration of a primary abstract model for finite-time heat engines, the efficiency at maximum power was proved to be universal for tight-coupling heat engines up to the linear term of the Carnot efficiency~\cite{vdbrk2005}.

\subsection{The second law of thermodynamics}
Four laws of thermodynamics lay the foundation of thermodynamics.
The zero law gives the condition of equilibrium, which enables us to design a thermometer. The first law, the law of energy conservation, is mathematically expressed as
\begin{equation}\Delta U=Q-W,\label{eq-engconserved}\end{equation}
where $\Delta U$, $Q$, $W$ are the internal energy increment of the concerned system, the heat absorbed from environment, the work output by the concerned system, respectively, in some process.

\begin{figure}[htp!]
\includegraphics[width=7cm]{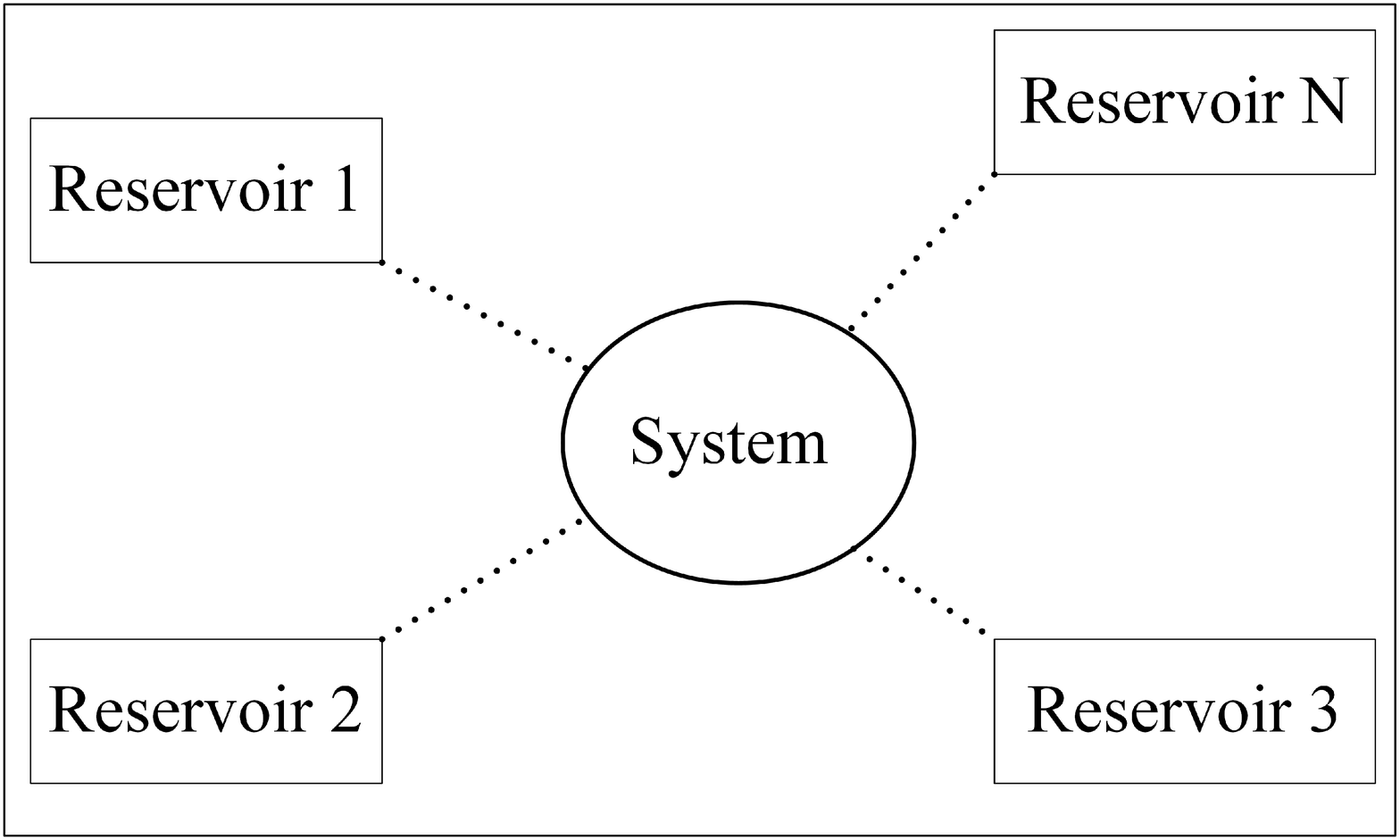}
\caption{A system in contact with $N$ reservoirs.}\label{fig_sys-env}
\end{figure}

The second law is the core of the four laws of thermodynamics. For an isolated system, the entropy never decreases. That is
\begin{equation}\Delta S_\mathrm{iso}\ge 0.\label{eq-endisol}\end{equation}
Now we consider a closed system in contact with $N$ reservoirs shown in Fig.~\ref{fig_sys-env}. The whole system (we call it universe) consisting of the system and reservoirs is an isolated system. According to (\ref{eq-endisol}), the entropy of the universe never decreases:
\begin{equation}\Delta S_{U}=\Delta S+\sum_{n=1}^N\Delta S_n^r\ge 0,\label{eq-entropyuniv}\end{equation}
where $\Delta S$ represents the entropy change of the system. $\Delta S_n^r=- Q_n/T_n$ represents the change of entropy for reservoir $n$. $T_n$ and $Q_n$ represent the temperature of reservoir $n$ and the heat absorbed from reservoir $n$ by the system, respectively.

The intrinsic entropy production in the system of interest is defined as $\Delta S^{(i)}\equiv\Delta S_U$. The entropy flow from reservoirs into the system is defined as:
\begin{equation}
\Delta S^{(e)}\equiv -\sum_{n=1}^N\Delta S_n^r=\sum_{n=1}^N \frac{Q_n}{T_n}\label{eq-entropyflow}\end{equation}
Then the entropy change (\ref{eq-entropyuniv}) of the system may be split into two parts:
\begin{equation}\Delta S= \Delta S^{(e)} + \Delta S^{(i)},\label{eq-sysentropy}\end{equation}
where the intrinsic entropy production $\Delta S^{(i)} \ge 0$. The equal sign holds for reversible processes while the greater than symbol holds for irreversible processes.

\begin{figure}[htp!]
\includegraphics[width=7cm]{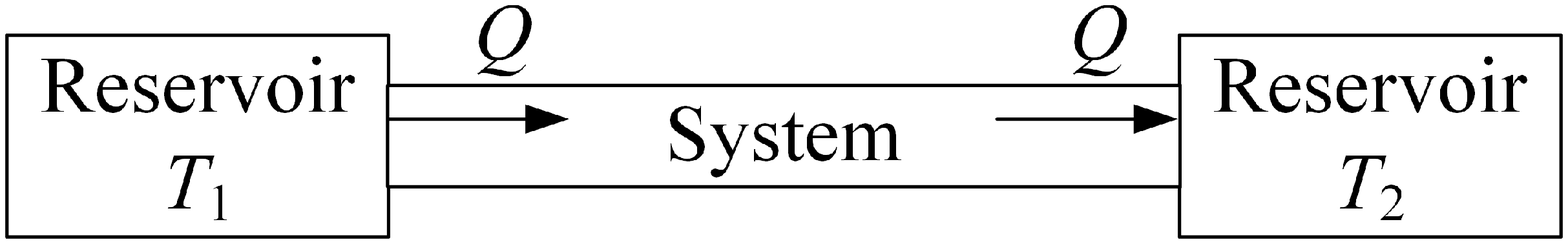}
\caption{Steady heat transfer.}\label{fig-heattrans}
\end{figure}

For an isolated system, $\Delta S^{(e)}=0$, Eq.~(\ref{eq-sysentropy}) implies $\Delta S\ge 0$, which is consistent to Eq.~(\ref{eq-endisol}). If we consider a closed system in contact with a reservoir at temperature $T$, $\Delta S^{(e)}=Q/T$. Then Eq.~(\ref{eq-sysentropy}) implies $\Delta S\ge Q/T$ since $\Delta S^{(i)}\ge 0$. This conclusion is in agreement with the Clausius inequality: when a closed system is in contact with a reservoir, the change of entropy is larger than the ratio of the absorbed heat to the temperature. As shown in Fig.~\ref{fig-heattrans}, we consider steady heat transfers between two reservoirs at different temperatures. In steady state, the heat $Q$ absorbed from reservoir 1 is the same as that released into reservoir 2, and vice verse. The entropy of the system is unchanged at the steady state, that is $\Delta S=0$. Since $\Delta S^{(i)}\ge 0$, we arrive at $Q(T_1-T_2)\ge 0$. The equal sign holds only for $T_1=T_2$. Thus $Q >0$ when $T_1>T_2$. This finding is consistent to the Clausius statement of second law: the heat spontaneously transfer from the high temperature end to the low temperature one.

\subsection{Linear irreversible thermodynamics}
The core concept in irreversible thermodynamics is the entropy production rate
\begin{equation}\sigma\equiv \lim_{\Delta t\rightarrow 0}\left({\Delta S^{(i)}}/{\Delta t}\right)\end{equation}
which is always non-negative. When the system is in equilibrium under certain external field parameters $\Lambda_k =\Lambda_{k0}$ ($k=1,2,\cdots$), the entropy production rate is vanishing. If $\Lambda_k \neq\Lambda_{k0}$, the system departs from equilibrium. Define a generalized thermodynamic force $X_k=\Lambda_k -\Lambda_{k0}$. The entropy production rate may be phenomenologically expressed in a canonical form~\cite{Prigoginebook}:
\begin{equation}\sigma\equiv \sum_k J_k X_k> 0, \label{eq-epr-can}\end{equation}
where $J_k$ is called the conjugated thermodynamic flux related to $X_k$.
The constitutive relation for response may be expressed as function $J_k = J_k (X_1, X_2 ,\cdots)$. Under the linear assumption for small departures,
The constitutive relation is simplified as: \begin{equation}J_k = \sum_l L_{kl} X_l,\label{eq-linearResponse}\end{equation}
where the coefficients satisfy the Onsager reciprocal relation~\cite{Onsager1931a,Casimir1945}:
\begin{equation}L_{kl}=L_{lk}\label{eq-reciprocal}\end{equation}
if the equations of microscopic motion keep invariant under time reversal. In particular, Eq.~(\ref{eq-epr-can}) implies that the matrix $[L_{kl}]$ is positively definite.

\subsection{Classic abstract model for heat engines\label{sec-classmodel}}
\begin{figure}[htp!]
\includegraphics[width=7cm]{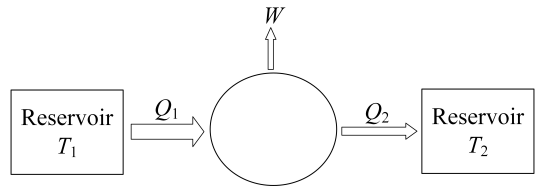}
\caption{Classic abstract model for heat engines.}\label{fig-engmod1}
\end{figure}

In the textbook of thermal physics, one can find the classic abstract model shown in Fig.~\ref{fig-engmod1} for heat engines.
The working substance absorbs heat $Q_1$ from the hot reservoir at temperature $T_1$, outputs work $W$, and releases heat $Q_2$ into the cold reservoir at temperature $T_2$.
For an autonomous engine in steady state, the entropy change of working substance $\Delta S=0$. Since $\Delta S^{(i)} \ge 0$, we derive
$\Delta S^{(e)}=Q_1/T_1-Q_2/T_2 \le 0$, which implies
\begin{equation}Q_2/Q_1\ge T_2/T_1.\label{eq-calssiceng1}\end{equation}
According to the first law of thermodynamics, we have the output work:
\begin{equation}W=Q_1-Q_2.\label{eq-1stlaw}\end{equation}
Thus the efficiency of the engine
\begin{equation}\eta=W/Q_1\le 1-T_2/T_1\equiv\eta_0.\label{eq-effcy}\end{equation}
For a cyclic engine, the above analysis still holds for each cycle. That is, the Carnot efficiency is the upper bound of the efficiencies for engines operating between two reservoirs at given temperatures.

\subsection{Primary abstract model for finite-time heat engines\label{sec-primarymodel}}
\begin{figure}[htp!]
\includegraphics[width=7cm]{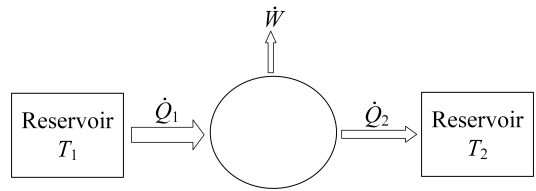}
\caption{Primary abstract model for finite-time heat engines.}\label{fig-engmod2}
\end{figure}

Now we turn to finite-time processes. The above abstract model can be modified slightly for finite-time heat engines. As shown in Fig.~\ref{fig-engmod2}, the absorbed heat $Q_1$, the work output $W$, and the released heat $Q_2$
are replaced with $\dot{Q}_1$ (heat absorbed from the hot reservoir per unit time), $\dot{W}$ (power output), $\dot{Q}_2$ (heat released into the cold reservoir per unit time), respectively.

Following Van den Broeck's ideal~\cite{vdbrk2005}, the mechanical flux $J_m$ is taken as the net rate of autonomous engines or the reciprocal of period of cyclic engines. The conjugated mechanical force is
\begin{equation}X_m=-\beta w,\label{eq-mechforce} \end{equation}
where $\beta$ is the energy scale corresponding to the temperature of hot or cold reservoirs. $w$ is the work output by the heat engine per elementary step for autonomous engines or per cycle for cyclic engines. Thus the power output may be expressed as
\begin{equation}P=\dot{W}=-\beta^{-1}J_mX_m. \label{eq-power} \end{equation}

The thermal flux is taken as $J_t=\dot{Q}_1$ or $\dot{Q}_2$. The conjugated thermodynamics force is
\begin{equation}
X_t=1/T_2-1/T_1.\label{eq-thermalforce} \end{equation}
The tight coupling condition may be expressed as
\begin{equation}J_t=\xi J_m, \label{eq-Jttightcp} \end{equation}
where $\xi$ can be regarded as the thermal energy element flowing through the work substance per elementary step for autonomous engines or per cycle for cyclic engines. The entropy production (\ref{eq-epr-can}) is transformed into
\begin{equation}\sigma= J_m X_m+J_tX_t=J_m A, \label{eq-epr-can2}\end{equation}
where
\begin{equation}A=X_m+\xi X_t \label{eq-affinity}\end{equation} is called affinity.

Now we consider the linear constitute relation for response,
$J_m=L_{mm}X_m+L_{mt}X_t$ and $J_t=L_{tm}X_m+L_{tt}X_t$. The tight coupling condition requires $\xi=L_{tm}/L_{mm}=L_{tt}/L_{mt}$.
Substituting the linear relation into Eq.~(\ref{eq-power}), and then optimizing power with respect to $X_m$, we obtain the optimal $X_m=-(\xi/2)X_t$.
Finally, we achieve the efficiency at maximum power:
\begin{equation}\eta_{mP}= \frac{P_{\max}}{\dot{Q}_1}=\frac{\eta_0}{2}+O(\eta_0^2). \label{eq-unieta1}\end{equation}
This equation is the key result in Ref.~\cite{vdbrk2005}, which implies that the efficiency at maximum power is universal for tight-coupling heat engines up to the linear term of the Carnot efficiency.

\section{Refined abstract model for finite-time heat engines\label{sec-refinedmodel}}
In the primary abstract model for finite-time heat engines mentioned above, the definition of $J_t$ and $\beta$ are ambiguous. What on earth should $J_t$ equal to $\dot{Q}_1$ or $\dot{Q}_2$? What on earth should $\beta$ equal to $1/T_1$ or $1/T_2$? Sheng and Tu introduced the concepts of weighted thermal flux and weighted reciprocal of temperature to eliminate the above ambiguity~\cite{ShengTuPRE14}. With the aid of these concepts, a refined abstract model for finite-time heat engines was established. We sketch the main idea in this section.

Consider the primary abstract model for finite-time heat engines shown in Fig.~\ref{fig-engmod2}. The first law of thermodynamics requires:
\begin{equation}\dot{Q}_{1}-\dot{Q}_{2}=\dot{W}.\label{eq-Econst}\end{equation}
Since the entropy variation of working substance is vanishing either for a cyclic engine in the whole cycle or an autonomous engine in the steady state, the entropy production rate $\sigma$ may be expressed as:
\begin{equation}\sigma =
\beta_2\dot{Q}_{2}-\beta_1\dot{Q}_{1},\label{eq-EnP0}\end{equation}
with $\beta_1=1/T_1$ and $\beta_2=1/T_2$.
Here and in the following the Boltzmann factor is set to $1$.

Introduce two nonnegative weighted numbers $s_1$ and $s_2$ such that $s_1+s_2 =1$.
Define the weighted reciprocal of temperature
\begin{equation}\beta\equiv s_{1}\beta_1+s_{2}/\beta_2\label{eq-wttemp}\end{equation}
and weighted thermal flux
\begin{equation}J_t\equiv  s_{1}\dot{Q}_{2}+ s_{2}\dot{Q}_{1}.\label{eq-wtthflux}\end{equation}
Then the entropy production rate may be transformed into a canonical form as (\ref{eq-epr-can2}).

\begin{figure}[htp!]
\includegraphics[width=7cm]{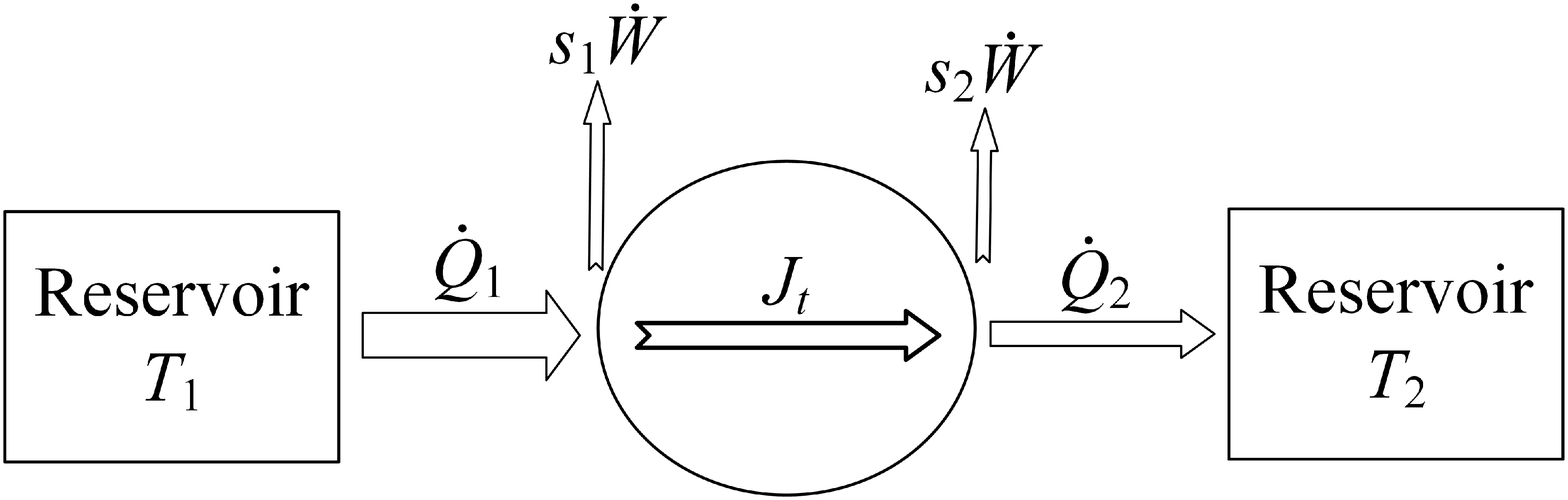}
\caption{Refined abstract model for finite-time heat engines.}\label{fig-engmod3}
\end{figure}

In particular, from first law (\ref{eq-Econst}) and definition (\ref{eq-wtthflux}) of weighted thermal flux, we arrive at a detailed law of energy conservation
\begin{equation}
\dot{Q}_{1}=J_t+s_1\dot{W},~\dot{Q}_{2}=J_t-s_2\dot{W}.\label{eq-detaileconst}\end{equation}
This detailed law of energy conservation enlightens us to construct a refined abstract model shown in Fig.~\ref{fig-engmod3} for finite-time heat engines.
In this model, the weighted thermal flux $J_t$ is interpreted as a heat flux flowing through the working substance. The working substance couples with the hot reservoir and absorbs heat per unit time $\dot{Q}_1$ from the hot reservoir. A certain amount of heat $s_1 \dot{W}$ is transformed into work output per unit time due to the coupling between the working substance and the hot reservoir. A thermal flux $J_t$ flows through the working substance, then an amount of heat $s_2 \dot{W}$ is transformed into the work output per unit time due to the coupling between the working substance and the cold reservoir. Finally, the working substance releases heat per unit time $\dot{Q}_2$ into the cold reservoir.

Since the mechanical flux is in the liner order for small thermodynamic force, Eq.~(\ref{eq-power}) implies that the leading term of $\dot{W}$ is a quadratic order term for small thermodynamic forces.
$\dot{Q}_{1}$ should contain at least the linear term for small relative temperature.
Combining the detailed law of energy conservation (\ref{eq-detaileconst}), we regard $J_t$ as the common leading term shared by $\dot{Q}_{1}$ and $\dot{Q}_{2}$. The parameter $s_1$ (or $s_2$) represents the fraction of power output occupied by the higher order term in $\dot{Q}_{1}$ (or $\dot{Q}_{2}$). We argue that $s_1$ (or $s_2$) depends on the degree of coupling between the working substance and the hot (or cold) reservoir. That is, the coupling between the working substance and the reservoirs is significant for finite-time heat engines although it is unimportant for the conventional quasi-static heat engines.

\begin{figure}[htp!]
\includegraphics[width=7cm]{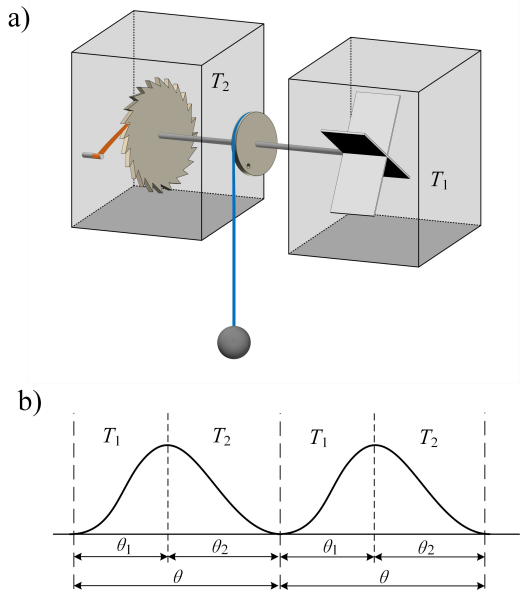}
\caption{Feynman ratchet: a) Cartoon of the Feynman ratchet; b) B\"uttiker-Landauer model.}\label{fig-ratchet}
\end{figure}

As an example, we consider the Feynman ratchet as shown in Fig.~\ref{fig-ratchet}a, which may be simplified as the the B\"uttiker-Landauer model~\cite{1Buttiker,1Landauer} shown in Fig.~\ref{fig-ratchet}b. A Brownian particle walks in a periodic potential with a fixed step size $\theta$, in contact with a hot (cold) reservoir at temperature $T_1$ ($T_2$) in the left (right) side of each energy barrier. The particle moves across each barrier from left to right and outputs work against a load $z$. The height of energy barrier is $\epsilon$. The width of potential in the left or right side of the barrier is denoted by $\theta_{1}$ or $\theta_{2}=\theta-\theta_{1}$, respectively.

In each forward step, the particle absorbs heat $\epsilon+z{\theta_1}$ from the hot reservoir. After outputting work $z\theta$ against the external load, the remained heat $\epsilon -z\theta _{2}$ will be released into the cold reservoir when the particle jumps over the barrier. The energy transformation in each backward step is exactly opposite of that in the forward step mentioned above.
The forward and backward jumping rates can be respectively expressed as
$R_{F} =r_{0}\mathrm{e}^{-\beta_1(\epsilon +z\theta _{1})}$ and $R_{B}=r_{0}\mathrm{e}^{-\beta_2(\epsilon -z\theta _{2})}$, where $r_{0}$ represents the bare rate constant with dimension of time$^{-1}$. The net rate may be defined as
\begin{equation}r\equiv R_{F}-R_{B}=r_{0}\left[ \mathrm{e}^{-\beta _{1}( \epsilon +z\theta _{1})
}-\mathrm{e}^{-\beta _{2}( \epsilon -z\theta _{2}) }\right].\label{eq-fluxFeym}\end{equation}
The net power output can be expressed as
\begin{equation}\dot{W}=z\theta r\label{eq-network}. \end{equation}
The heat absorbed from the hot reservoir and that released into the cold reservoir per unit time can be respectively expressed as
\begin{equation}\dot{Q}_{1}=(\epsilon+z\theta _{1})r=\epsilon r+ (\theta _{1}/\theta)\dot{W}\label{eq-Qhot}, \end{equation}
and
\begin{equation}\dot{Q}_{2}=(\epsilon -z\theta _{2})r=\epsilon r- (\theta _{2}/\theta)\dot{W}\label{eq-Qcold}. \end{equation}
Considering the above two equations and the detailed law of energy conservation (\ref{eq-detaileconst}), we can straightforwardly write out the weighted thermal flux
\begin{equation}J_t=\epsilon r\label{eq-JtFeym},\end{equation}
and the weighted parameters
\begin{equation}s_1=\theta _{1}/\theta,~s_2=\theta _{2}/\theta\label{eq-wtnumFeym}.\end{equation}
Obviously, the above weighted parameters reflect the relative coupling regions between the Brownian particle and the hot reservoir or the cold one in Fig.~\ref{fig-ratchet}b, which indeed depend on the degree of coupling between the working substance and the hot (or cold) reservoir.
Furthermore, the mechanical flux may be expressed
\begin{equation}J_{m}\equiv r=r_{0}\textrm{e}^{-\bar{\beta}\epsilon}A\left[1+\frac{\lambda}{2}(A-\epsilon X_{t})\right] +O(A^{3}, X_{t}^{3}),\label{Eq-Feynman-Jm}\end{equation}
with $\bar{\beta}=(1/T_{1}+1/T_{2})/2$. $\lambda\equiv s_1-s_2$ represents the asymmetry degree of coupling between the working substance and two reservoirs.

In fact, Eq.~(\ref{Eq-Feynman-Jm}) is a specific form of generic expression of constitutive relation for nonlinear response.
For tight-coupling heat engines, Sheng and Tu phenomenologically derived a generic constitutive relation for nonlinear response accurate up to the quadratic order terms from the stalling condition and the symmetry argument~\cite{ShengTuPRE15}. This constitutive relation may be expressed as
\begin{equation}J_{m}=LA\left[1+\alpha \lambda (A+uX_{t})\right]+O(A^{3},X_{t}^{3}),\label{Eq-solution-Jm2}\end{equation}
where $L$, $\alpha$ and $u$ are model-dependent coefficients. Considering the homotypic nature of heat engines in literature, Sheng and Tu further proved two facts: i) $\alpha$ is independent of $\lambda$; ii) $u=0$ or $-\xi$ for cyclic or autonomous heat engines, respectively~\cite{ShengTuNJP15}.

Using relation (\ref{Eq-solution-Jm2}), one can achieve the efficiency at maximum power~\cite{ShengTuPRE15}:
\begin{equation} \eta_{mP}=\frac{1}{2}\eta_{0}+\frac{1}{8}\eta_{0}^{2}+\frac{\lambda(1-\alpha\beta\xi)}{8}\eta_{0}^{2}+O(\eta_{0}^{3}),\label{Eq-solution-eta}\end{equation}
from which we see that the third term vanishes when $\lambda (1-\alpha\beta\xi)= O(\eta_0)$. This condition may be further expressed as~\cite{ShengTuPRE15}:
\begin{equation}\lambda =0+O(\eta_0)~\text{or}~\alpha\beta\xi=1+O(\eta_{0}).\label{Eq-solution-condition}\end{equation}
That is, we obtain the necessary and sufficient condition for the universal prefactor 1/8 of the quadratic term in (\ref{eq-univ}).
The former one is called symmetry-coupling condition while the latter one is called energy-matching condition. It is shown that both the Curzon-Ahlborn endoreversible engine and the Feynman ratchet satisfy the energy-matching condition~\cite{ShengTuPRE15}. This is the underlying reason why the Curzon-Ahlborn endoreversible engine and the Feynman ratchet recover the universal efficiency at maximum power regardless of any symmetry. Thus the apparent paradox mentioned in the beginning of this review is resolved.

\section{Detailed models as special cases of refined abstract model\label{sec-appl}}
The refined abstract model is quite general for finite-time heat engines. We have shown that the Feynman ratchet may be mapped into the refined abstract model in the above section. Here, we will demonstrate that other typical finite-time heat engines in literature can also be mapped into the refined abstract model~\cite{ShengTuPRE14,ShengTuPRE15}.

\subsection{Low-dissipation engine\label{sec-lowdispat}}
\begin{figure}[htp!]
\includegraphics[width=7cm]{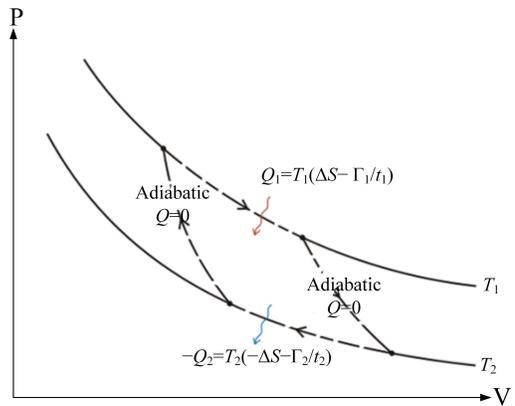}
\caption{Thermodynamic cycle of low-dissipation engines.}\label{fig-lowdiseng}
\end{figure}

A low-dissipation engine~\cite{Esposito2010} undergoes a thermodynamic cycle consisting of two ``isothermal" and two adiabatic processes shown in Fig.~\ref{fig-lowdiseng}. The word ``isothermal" merely indicates that the heat engine is in contact with a thermal reservoir at constant temperature.

The heat absorbed from the hot reservoir and that released into the cold one satisfy the following low-dissipation assumption~\cite{Esposito2010}:
\begin{equation}Q_{1} =T_{1}( \Delta S-\Gamma_{1}/t_{1}),~
-Q_{2} =T_{2}( -\Delta S-\Gamma_{2}/t_{2}),\label{eq-lowdisassp}\end{equation}
with two dissipation coefficients $\Gamma_{1}$ and $\Gamma_{2}$.
$t_1$ ($t_2$) is the time for completing the ``isothermal" expansion (compression). Assume that the time for completing the adiabatic processes is negligible relative to $t_1$ and $t_2$. So the period of the whole cycle is $t_0=t_1+t_2$.

The low-dissipation engine may be mapped into the refined abstract model~\cite{ShengTuPRE14}. The weighted numbers are
\begin{equation}s_{1}=\frac{T_{1}\bar\Gamma_{1}}{T_{1}\bar\Gamma_{1}+T_{2}\bar\Gamma_{2}},~s_{2}=\frac{T_{2}\bar\Gamma_{2}}{T_{1}\bar\Gamma_{1}+T_{2}\bar\Gamma _{2}};\label{eq-cpparm}\end{equation}
with parameters $\bar\Gamma_{1}\equiv \Gamma_{1}t_{0}/t_{1}$ and $\bar\Gamma_{2}\equiv \Gamma_{2}t_{0}/t_{2}$.
Interestingly, the constitutive relation is linear form~\cite{ShengTuPRE15}:
\begin{equation}J_m =\frac{1}{\bar\Gamma _{1}+\bar\Gamma _{2}}A,\label{eq-Jmloweng}\end{equation}
which is a special form of generic constitutive relation (\ref{Eq-solution-Jm2}) with $\alpha=0$ and vanishing higher order terms.
In addition, the minimally nonlinear irreversible heat engines proposed by Izumida and Okuda~\cite{Izumida2012}, and the autonomous thermoelectric generators investigated by Apertet \emph{et al.}~\cite{ApertetPRE13} can also be mapped into the refined abstract model. Surprisingly, the constitutive relations for these engines are also linear form~\cite{ShengTuPRE15}.

Obviously, $\alpha =0$ implies that the energy-matching condition in Eq.~(\ref{Eq-solution-condition}) cannot be satisfied. Thus, this kind of engines take the universal efficiency at maximum power if and only if the symmetry-coupling condition is satisfied. Note that the Schmiedl-Seifert engine~\cite{Schmiedl2008} is equivalent to a symmetric low-dissipation engine, which explains the reason why the Schmiedl-Seifert efficiency has the universal form up to the quadratic term of $\eta_0$.

\subsection{Curzon-Ahlborn heat engine\label{Sec-CAengine}}

The Curzon-Ahlborn endoreversible heat engine~\cite{Curzon1975} undergoes a cycle consisting of two ``isothermal" processes and two adiabatic processes. The effective temperature of working substance is assumed to be $T_{1e}$ ($T_{2e}$) when it is in contact with a hot (cold) reservoir. The engine seems to reversibly operate between two reservoirs at temperatures $T_{1e}$, $T_{2e}$. This is the endoreversible assumption~\cite{Curzon1975} which is mathematically expressed as
\begin{equation}Q_1/T_{1e}=Q_1/T_{2e}\equiv\Delta S.\label{eq-endorevassum}\end{equation}

The heat transferred from the hot reservoir to the working substance abides by the Newton-type heat transfer law
\begin{equation}\label{eq-transQh}
Q_{1}=\kappa_{1}(T_{1}-T_{1e})t_{1},\end{equation}
where $\kappa_{1}$ and $t_1$ are the thermal conductivity and the time interval of ``isothermal" expansion, respectively.
Similarly, the heat transferred from  the working substance to the cold reservoir may be expressed as
\begin{equation}\label{eq-transQc}Q_{2}=\kappa_{2}(T_{2e}-T_{2})t_{2},\end{equation}
where $\kappa_{2}$ and $t_2$ are the thermal conductivity and the time interval of ``isothermal" compression, respectively. The period ($t_{0}$) for completing the whole cycle is assumed to be proportional to $t_{1}+t_{2}$.

The Curzon-Ahlborn endoreversible heat engine can be mapped into the refined abstract model~\cite{ShengTuPRE15}. The weighted numbers are as follows:
\begin{equation}s_{1}=\frac{T_{1}\gamma_{2}}{T_{1}\gamma_{2}+T_{2}\gamma_{1}},~s_{2}=\frac{T_{2}\gamma_{1}}{T_{1}\gamma_{2}+T_{2}\gamma_{1}}\label{eq-CA4}\end{equation}
with parameters $\bar\Gamma_{1}\equiv \Gamma_{1}t_{0}/t_{1}$ and $\bar\Gamma_{2}\equiv \Gamma_{2}t_{0}/t_{2}$.
It was found that the constitutive relation is also a special form of generic expression (\ref{Eq-solution-Jm2}) with model-dependent parameters $L=\gamma_{1}\gamma_{2}/(\gamma_{1}+\gamma_{2})\Delta S^{2}$, $\alpha=1/\Delta S$, $u=0$ and $\xi=T_{1}T_{2}\beta \Delta S$~\cite{ShengTuPRE15}. It is easy to verify $\alpha\beta\xi=1+O(\eta_{0})$, which conforms with the energy-matching condition in (\ref{Eq-solution-condition}). This is the underlying reason why the Curzon-Ahlborn endoreversible heat engine recovers the universal efficiency at maximum power regardless of any symmetry.

Chen and Yan also investigated a revised Curzon-Ahlborn endoreversible heat engine~\cite{Chen1989} which is exactly the same as the original Curzon-Ahlborn engine except for the heat transfer law in two ``isothermal" processes:\begin{equation}Q_{1}=\kappa_{1} (T_{1e}^{-1}-T_{1}^{-1})t_{1},~Q_{2}=\kappa_{2} (T_{2}^{-1}-T_{2e}^{-1})t_{2}.\label{eq-rvCAQhc}\end{equation}
The revised Curzon-Ahlborn heat engine may also be mapped into the refined abstract model~\cite{ShengTuPRE15}. The weighted numbers are as follows:
\begin{equation}s_{1}=\frac{T_{1}^{3}/\gamma_{1}}{T_{1}^{3}/\gamma_{1}+T_{2}^{3}/\gamma_{2}},~s_{2}=\frac{T_{2}^{3}/\gamma_{2}}{T_{1}^{3}/\gamma_{1}+T_{2}^{3}/\gamma_{2}}\label{revCA-shsc}\end{equation}
with $\gamma_{1}\equiv\kappa_{1}t_{1}/t_{0}$, $\gamma_{2}\equiv\kappa_{2}t_{2}/t_{0}$.
The constitutive relation is also a special form of generic expression (\ref{Eq-solution-Jm2}) with model-dependent parameters $L=1/(T_{1}^{2}/\gamma_{1}+T_{2}^{2}/\gamma_{2})\Delta S^{2}$, $\alpha =2/\Delta S$, $u=0$ and $\xi=T_{1}T_{2}\beta \Delta S$~\cite{ShengTuPRE15}. It is easy to prove $\alpha\beta\xi=2+O(\eta_{0})$, which implies the energy-matching condition in Eq.~(\ref{Eq-solution-condition}) cannot be satisfied by this engine. Thus, the revised Curzon-Ahlborn heat engine takes the universal efficiency at maximum power if and only if the symmetry-coupling condition is satisfied. This conclusion is consistent with equation (31) in Ref.~\cite{Chen1989}.

\subsection{Single-level quantum-dot heat engine\label{sec-quantum}}
\begin{figure}[htp!]
\includegraphics[width=7cm]{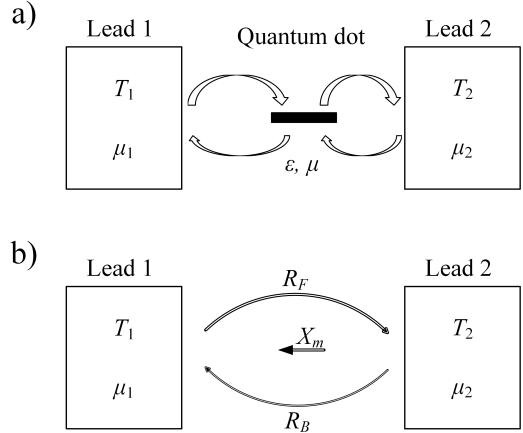}
\caption{Single-level quantum dot: a) detailed model; b) simplified model.}\label{fig-quandot}
\end{figure}

As shown in Fig.~\ref{fig-quandot}a, a single-level quantum-dot heat engine~\cite{Esposito2009a} consists of three parts: a hot lead with temperature $T_{1}$ and chemical potential $\mu_{1}$; a cold lead with temperature $T_{2}$ ($T_{2}<T_{1}$) and chemical potential $\mu_{2}$ ($\mu_{2}>\mu_{1}$); a single-level quantum dot with energy level $\varepsilon$ ($\varepsilon>\mu_{2}$) and chemical potential $\mu$ ($\mu_{2}>\mu>\mu_{1}$), which located between the two leads. In the forward process, an electron jumps from the hot lead 1 to the cold lead 2 via the quantum dot. In the backward process, an electron jumps from the cold lead 2 to the hot lead 1 via the quantum dot. The jumping rates are preset parameters.

In the steady state, the single-level quantum-dot heat engine can be simplified as Fig.~\ref{fig-quandot}b where the generalized mechanical force is $X_m=-\beta(\mu_2-\mu_1)$. The forward electronic flow and the backward one are
\begin{equation}R_{F}=\frac{ I_0}{\mathrm{e}^{(\varepsilon-\mu_{1})/T_{1}}+1}~\mathrm{and}~R_{B}=\frac{I_0 }{\mathrm{e}^{(\varepsilon-\mu_{2})/T_{2}}+1},\label{Eq-quantum-IFB}\end{equation}
respectively. Here $I_0$ is a temperature-independent coefficient.

This model can be further mapped into the refined abstract model~\cite{ShengTuPRE15}. The weighted numbers are as follows:
\begin{equation}s_{1}=\frac{\mu-\mu_{1}}{\mu_2-\mu_1},~s_{2}=\frac{\mu_{2}-\mu}{\mu_2-\mu_1}.\label{Eq-quantum-scsh}\end{equation}
It was found that the constitutive relation is also a special form of generic expression (\ref{Eq-solution-Jm2}) with model-dependent parameters $\xi=\varepsilon-\mu$, $L=I_0/4\cosh^{2}(\bar{\beta}\xi/2)$, $\alpha=(1/2)\tanh(\bar{\beta}\xi/2)$ and $u=-\xi$. When the engine operates at maximum power, we can verify $\alpha\beta\xi=1+O(\eta_{0})$~\cite{ShengTuPRE15}. Thus, this engine always recovers the universal efficiency when operating at maximum power, which is consistent with the results in Ref.~\cite{Esposito2009a}.

\section{Summary and outlook\label{sec-summary}}
We have sketched three abstract models for heat engines in the above contents which include a classic abstract model for heat engines, a primary abstract model for finite-time heat engines, and a refined abstract model for finite-time heat engines.
From the pedagogical point of view, the refined abstract model for finite-time heat engines corresponds to the classic abstract model for heat engines in thermal physics.
From the practical point of view, this refined abstract model has not only manifested applicable significance in proving the conjecture of universal efficiency at maximum power, but also provided a unified description to typical finite-time heat engines in literature.
In the future, we expect that this refined abstract model will be utilized in discussing the following issues.

i) Efficiency fluctuations. Microscale heat engines were constructed in several recent experiments. Different from macroscopic engines, the efficiency and power of microscale heat engines exists relatively large fluctuations. Blickle and Bechinger realized a micrometre-sized stochastic heat engine with a single colloidal particle subject to a time-dependent optical laser trap~\cite{BechingerNP12}. This is the first time to realize the Schmiedl-Seifert stochastic engines in experiment. They did not confirm theoretical predictions by Schmiedl and Seifert, but also observed the fluctuation of efficiency. Abah~\emph{et al.} designed a single-ion quantum Otto engine by confining an ion in a linear Paul trap~\cite{Lutz2012PRL}. This engine can run at maximum power in a wide range of temperatures. Their experimental results also reveal fluctuating behaviors of efficiency. These experiments enlighten theoretical investigations on efficiency fluctuations. Verley~\emph{et al.} depicted the universal features of efficiency fluctuations by using the fluctuation theorem~\cite{EspositoNC2014}. The most surprising finding is that the Carnot efficiency is the least likely in the long-time limit. Generally, the reversible and the least likely efficiency coincide for time-symmetric driving~\cite{VerleyPRE2014}. Jiang~\emph{et al.} revealed the universal behavior of stochastic efficiency for time-reversal symmetry-broken engines in the Gaussian approximation by using large deviation theory~\cite{JiangJHPRL2015}. Park~\emph{et al.} obtained the large deviation function of the probability distribution for the stochastic efficiency for a linear Brownian heat-engine model in the overdamped limit~\cite{NohPRE2016}. If we consider the finite-time  efficiency and power fluctuations, the refined abstract model may be taken as a start point. The fluctuations of weighted thermal flux $J_t$ determine the behaviors of efficiency and power fluctuations.

ii) Trade-off between power and efficiency. A heat engine usually outputs vanishing power when it achieves maximum efficiency (the Carnot efficiency). Its efficiency is quite smaller than the Carnot efficiency when it operates at maximum power. That is, there exists a trade-off between power and efficiency. This trade-off should be essentially related to the thermodynamic uncertainty relation proposed by Barato and Seifert~\cite{BaratoPRL2015}. Rather than the issue of efficiency at maximum power, Allahverdyan \emph{et al.} asked a good question~\cite{AllahverdyanPRL3}: What is the maximum efficiency at a given power? Holubec and Ryabov semi-analytically derived the universal bounds on maximum efficiency at a given power for low-dissipation engines~\cite{HolubecJSM2016}. This is the first exact relation on trade-off between power and efficiency. Ma \emph{et al.} provided a fully analytical proof on this trade-off relation for low-dissipation engines~\cite{MaYHPRE2018}. Holubec and Ryabov also discussed the trade-off relation for steady-state heat engines on the basis of linear irreversible thermodynamics~\cite{HolubecPRE2016}. Proesmans \emph{et al.} proposed a power-efficiency-dissipation relation within the framework of linear irreversible thermodynamics~\cite{ProesmansPRL2016}. Iyyappan and Ponmurugan investigated the power-efficiency-dissipation relation for minimally nonlinear irreversible heat engines~\cite{IyyappanPRE2018}. Shiraishi \emph{et al.} generalized their discussions and derived a lower bound for dissipation from which they obtained a universal trade-off relation between power and efficiency for heat engines~\cite{ShiraishiPRL2016}. Pietzonka and Seifert derived a universal trade-off between power, efficiency, and constancy for steady-state engines~\cite{PietzonkaPRL2018}. We may investigate the trade-off between power and efficiency in detail using the refined abstract model. We expect that the variation and average of $J_t$ have a similar thermodynamic uncertainty relation. This relation may imply a trade-off between power and efficiency.

iii) Thermodynamic cycles controlled with shortcuts. In finite-time thermodynamics, the discussion on adiabatic processes is usually ignored. Generally speaking, the adiabaticity requires a slowly enough controlling. Are finite-time adiabatic processes possible? Shortcut to adiabaticity~\cite{AoP2000,Rice2003,Berry2009,ChenPRL10,Jarzynski40101,Campo100502,DJdCAX14,MugaRMP2019} is a novel concept to realize finite-time adiabatic processes. Deng \emph{et al.} found that shortcuts to adiabaticity enhance the performance of heat engines~\cite{DengJ4182}. Tu introduced shortcuts to adiabaticity in the Schmiedl-Seifert model with the consideration of the inertial effect of Brownian particles~\cite{TuPRE2014}. Surprisingly, the efficiency at maximum power of this microscopically stochastic model is found to exactly equal to the Curzon-Ahlborn efficiency. Abah and Lutz demonstrated that shortcuts to adiabaticity simultaneously enhance the efficiency and power for quantum engines~\cite{Abah2018}. Recently, a  strategy of shortcuts to isothermality was proposed to accelerate isothermal processes~\cite{LiTuPRE2017}. This strategy has been realized by Albay \emph{et al.} in experiments~\cite{Albay2019,Albay2020}. It is worth investigating a thermodynamic cycle with two shortcuts to isothermality and two shortcuts to adiabaticity~\cite{PancottiPRX2020,NakamuraPRE2020}. This model may also be mapped into the
refined abstract model. It is interesting to discuss the efficiency and power fluctuations, the trade-off relations between power and efficiency of this kind of engines.

v) Universal efficiency at maximum $\Omega$-criterion. The optimization of heat devices based on a compromise between energy benefits and losses has been discussed frequently. In particular, Calvo Hern\'{a}ndez and his coworkers proposed the so-called $\Omega$-criterion representing this compromise~\cite{achPRE037102}. The efficiencies at optimal $\Omega$-criterion for the Curzon-Ahlborn endoreversible engine, the Schmiedl-Seifert engine, and the Feynman ratchet, the low-dispassion engine were investigated~\cite{achPRE051101,achPRE012105}. These efficiencies also reveal a universality, which has been demonstrated by Zhang \emph{et al.} with a tight-coupling model of particle-transports in the presence of left-right symmetry~\cite{ChenjcPRE032152}. It is straightforward to discuss the universality of efficiency at optimal $\Omega$-criterion with the consideration of the refined abstract model. We expect to achieve the energy-matching condition as we found for the universality of efficiency at maximum power.

vi) Non-homotypic engines. It is found that typical heat engines in literature exhibit a kind of homotypy~\cite{ShengTuNJP15}. The heat exchanges between a cyclic heat engine and its two reservoirs are dual under the joint transformation of parity inversion and time-reversal operation. Similarly, the forward and backward flows in an autonomous heat engine are also dual under the parity inversion. In other words, there exists hidden symmetries for homotypic heat engines. If breaking these hidden symmetries, one will construct many models of non-homotypic engines. The study on non-homotypic engines is quite rare. The only reference is contributed by Zhao and Tu who investigated nonlinear constitutive relation and efficiency at maximum power of non-homotypic engines~\cite{ZhaoTuBNU2016}. There is plenty of scene on the study of non-homotypic engines. On may construct refined abstract model of non-homotypic engines and then investigate the universality of efficiency at maximum power or other criterions.

vii) Active heat engines. All engines mentioned above are passive objects. If an engine is constituted by active particles as working substance or thermal reservoirs, its behaviors should be quite different from the passive engines since this active engine is a naturally nonequilibrium system. Krishnamurthy \emph{et al.} experimentally realized a micrometre-sized Stirling engine operating between active bacterial reservoirs~\cite{Krishnamurthy16}. They observed that the non-Gaussianity of reservoir noise largely influences the performance of engines.
Mart\'{\i}nez \emph{et al.} discussed the work extraction from colloidal engines operating between active bacterial reservoirs which could significantly boost the performance of engines~\cite{RicasoftM2017}. Pietzonka \emph{et al.} investigated the energetics of an autonomous engines driven by active particles \cite{PietzonkaPRX04103}. They found that the interaction with passive objects can enhance the extracted power per active particle.
Ekeh \emph{et al.} designed thermodynamic cycles with active particles and investigated the power and efficiency of this kind of active engines~\cite{Catesactive}. They found that the engines may simultaneously reach maximum power and efficiency. Kumari \emph{et al.} investigated the Stirling engine by using an self-propelling particle (as active working substance) in contact with two thermal reservoirs and showed that the active properties enhance the performance of the engine~\cite{LahiriPRE2020}. Lee \emph{et al.} considered a solvable linear model of engines with active reservoirs and defined proper temperatures of active reservoirs, base on which they demonstrated that the efficiency can overcome the Carnot bound~\cite{LeeParkPRE032116}. Is there a generic abstract model for active heat engines? Can we construct an abstract model for finite-time active heat engines? If we can do that, we are able to explore universal behaviors for active heat engines, such as efficiency, power, the trade-off between efficiency and power, the fluctuations of efficiency and power.

\section*{Acknowledgement}
The author is grateful for financial supports from the National Natural Science Foundation of China (Grant No. 11975050 and No. 11675017).

\end{document}